\documentclass[a4paper,11pt]{article}

\usepackage{jheppub} 
\usepackage[T1]{fontenc} 
\usepackage{lmodern} 
\usepackage{amsmath,mathrsfs}

\def\cN{\mathcal{N}}

\def\cW{\mathcal{W}}
\def\sW{\mathscr{W}}
\def\cO{\mathcal{O}}

\usepackage{slashed}

\def\){\right)}
\def\({\left( }
\def\]{\right] }
\def\[{\left[ }

\def\beq#1\eeq{\begin{align}#1\end{align}}

\newcommand{\bea}{\begin{eqnarray}}
\newcommand{\eea}{\end{eqnarray}}
\newcommand{\ra}{\rightarrow}

\newcommand{\nn}{\nonumber}
\newcommand{\tz}{{\tilde z}}

\title{
The Hamilton-Jacobi Equation and \\Holographic Renormalization Group Flows on Sphere}

\author{Nakwoo Kim and Se-Jin Kim}

\affiliation{Department of Physics and Research Institute of Basic Science,
	Kyung Hee University, \\ 26 Kyungheedae-ro, Dongdaemun-gu, Seoul 02447, Republic of Korea}
\emailAdd{nkim@khu.ac.kr}
\emailAdd{power817@khu.ac.kr}

\abstract{We study the Hamilton-Jacobi formulation of effective mechanical actions associated with holographic renormalization group flows when the field theory is put on the sphere and mass terms are turned on. Although the system is supersymmetric and it is described by a superpotential, Hamilton's characteristic function is not readily given by the superpotential when the boundary of AdS is curved. We propose a method to construct the solution as a series expansion in scalar field degrees of freedom. The coefficients are functions of the warp factor to be determined by a differential equation one obtains when the ansatz is substituted into the Hamilton-Jacobi equation. We also show how the solution can be derived from the BPS equations without having to solve differential equations. The characteristic function readily provides information on holographic counterterms which cancel divergences of the on-shell action near the boundary of AdS.}

\begin{document} 
\maketitle
\flushbottom

\section{Introduction}
\label{sec:intro}
A study in holography \cite{Maldacena:1997re}, in simple settings where the boundary is maximally symmetric and the bulk configuration also respects the symmetry, requires one to solve a system of coupled non-linear ordinary differential equations. After an appropriate procedure of regularization and renormalization \cite{deHaro:2000vlm,Bianchi:2001kw}, the supergravity action evaluated using the solution will then give the gravity-side computation of the path integral for dual gauge field theory.

In the terminology of soliton physics, such problems are classified as domain walls, with co-dimension one. In this paper we consider supersymmetric configurations with associated BPS equations, having comparison against the large-$N$ field theory computations in mind. Such first-order differential equations can be derived from Killing spinor equations, which are obtained by setting the fermionic variation of a bosonic configuration to zero.

Although the detail depends on spacetime dimensionality and the number of supersymmetries, the Lagrangian and its transformation rules are usually described in terms of a superpotential (and also the K\"ahler potential which determines the kinetic terms). The same applies to BPS equations naturally. 
 
There exists an elementary technique to derive BPS equations independently of supersymmetry: Starting from the Lagrangian with the solution ansatz substituted, one completes the square and show that when a certain first-order relation is satisfied the Lagrangian is reduced to a total-derivative. It proves that the BPS equations thus derived imply the field equations are satisfied, and as a byproduct we also have the on-shell action. Calculating the on-shell action is important in a number of applications. It gives the tunneling amplitude as an instanton in quantum field theory, while in the context of AdS/CFT correspondence the gauge theory path integral is given as the supergravity action. 

Let us illustrate the ``completing the square'' prescription using a simple model. Consider the following Lagrangian\footnote{This system can be derived as a static configuration of a supersymmetric field theory. The parameter $t$ is thus originally one of spatial coordinates. Alternatively, it can be also interpreted as instanton equation.} whose potential is given in terms of a ``superpotential'' $W$,
\beq
 L &= \frac{1}{2} g_{ab} \dot x^a \dot x^b + \frac{1}{2} g^{ab}
\frac{\partial W}{\partial x^a}\frac{\partial W}{\partial x^b}
\nn\\
&= \frac{1}{2} g_{ab}\left( \dot x^a \mp \frac{\partial W}{\partial x^c} g^{ac} \right)
\left( \dot x^b \mp \frac{\partial W}{\partial x^d} g^{bd} \right)  
\pm \frac{dW}{dt} . 
\eeq
From the above simple manipulation we have derived a BPS equation $\dot x^a = \pm g^{ab} ({\partial W}/{\partial x^b})$, and that the on-shell action is $S=\int dt L =\pm W$.

In the above example, the superpotential $W$ has an alternative interpretation. Namely, it provides a solution to Hamilton-Jacobi (HJ) equation with zero energy. Recall that, given a Hamiltonian $H(x^a,p_a)$, the HJ equation for Hamilton's characteristic function $W$ is given as 
\beq
H ( x^a, \frac{\partial W}{\partial x^a}) = E . 
\eeq
It is easily seen that the superpotential $W$ is actually Hamilton's charateristic function for the BPS system given above. It is also the case in a number of variations.

However, in a certain class of BPS systems the integration of HJ equation, or ``completing the square'' is not as straightforward as above, making the calculation of on-shell action non-trivial. It is this observation which motivated the current work. BPS systems with such a rather mysterious property are encountered in various settings, for instance from the study of multi-charge solitons in a supersymmetric field theory \cite{Kim:2007aa}. But in this paper we will specifically study curved BPS domain walls from supergravity. Earlier works can be found {\it e.g.} in \cite{Behrndt:2001mx,LopesCardoso:2001rt,Chamseddine:2001ga,LopesCardoso:2002ec} where the main focus was on how to derive the BPS equations for different supergravity models in different dimensions. 

From a contemporary perspective, curved BPS domain walls appear mainly in two different contexts of supergravity. One is as Janus solutions which have AdS space as the slice \cite{Bobev:2013yra,Suh:2011xc,Janik:2015oja,Pilch:2015dwa,Karndumri:2016tpf,Karndumri:2017bqi,Gutperle:2017nwo,Suh:2018nmp,Gutperle:2018fea,Kim:2020unz,Karndumri:2020led,Bobev:2020fon,Gutperle:2020gez}, and the other is holographic renormalization group flow with mass deformation dual to gauge field theory defined on the sphere \cite{Freedman:2013ryh,Bobev:2013cja,Bobev:2016nua,Gutperle:2018axv,Bobev:2018wbt}, which is of our interest in this paper. Note that, applying the Hamilton-Jacobi formalism to general relativity problems, in particular to holographic renormalization and AdS black holes where the radial coordinate is interpreted as ``time'' has a long history \cite{Parry:1993mw,Papadimitriou:2004ap,Batrachenko:2004fd,Papadimitriou:2006dr,Papadimitriou:2007sj,Papadimitriou:2010as,Gnecchi:2014cqa,Lindgren:2015lia,Elvang:2016tzz,Papadimitriou:2017kzw,Klemm:2017pxv,Cabo-Bizet:2017xdr,Castro:2018ffi}. In most of these works when the model of interest is supersymmetric the HJ equation is readily integrated by the superpotential. Models of sphere-sliced domain wall which allows an exact solution to HJ equation can be found {\it e.g.} in \cite{Papadimitriou:2007sj}.

Let us now explain why we are interested in sphere-sliced domain walls in supergravity.
The advent of supersymmetric localization technique \cite{Pestun:2016zxk} enabled precision tests of holography. The derivation of $N^{3/2}$ scaling of degrees of freedom for ABJM model on $S^3$ \cite{Aharony:2008ug,Drukker:2010nc} with the right coefficient for free energy was a particularly impressive feat for string duality and supersymmetric quantum field theory technology. 
As is well-known, conformal invariance does not predicate supersymmetric localization.  That said, the simplest tweak one can think of is to consider less-supersymmetric cases with arbitrary R-charge assignments for chiral multiplets, or equivalently turning on real mass which explicitly breaks conformal symmetry \cite{Herzog:2010hf,Cheon:2011vi,Martelli:2011qj,Jafferis:2011zi,Guarino:2015jca}. The large-$N$ results can be still obtained in closed form as a function of trial R-charges which then needs to be extremized according to the F-theorem \cite{Jafferis:2010un}. 

The dual procedure on gravity side  is comparatively more non-trivial and it is in general hard to find the holographic free energy as on-shell supergravity action in closed form. Supergravity BPS equations for mass deformed theories on the sphere have been derived for several models \cite{Freedman:2013ryh,Bobev:2013cja,Bobev:2016nua,Gutperle:2018axv,Bobev:2018wbt}. To construct them, one switches to Euclidean signature, identify the scalar fields in supergravity which are dual to supersymmetric mass terms in field theory, and derive the conditions on bosonic fields when we demand a non-trivial solution to Killing spinor equations. 

The dual of ABJM model, where one studies the Einstein-scalar sector of STU gauged supergravity in $D=4$, is the only case which allows BPS solutions in closed form. For other cases, the authors of \cite{Bobev:2013cja,Bobev:2016nua,Gutperle:2018axv,Bobev:2018wbt} relied on numerical analysis to establish relations between UV expansion coefficients which are dual to the source and vacuum expectation values of real mass operators in field theory and determine the renormalized supergravity action. Related works can be found in {\it e.g.} \cite{Balasubramanian:2013esa,Bigazzi:2013xia,Karch:2015kfa,Kol:2016ucd,Kim:2018sdw,Bobev:2018hbq,Bobev:2018eer,Bobev:2019wnf}.

More recently, the present authors proposed and pursued a perturbative prescription which allows one to obtain the holographic action as a power series in UV expansion parameters. 
The conjectures based on numerical results are partly confirmed for $\cN=2^*$ and $\cN=1^*$ deformations of $\cN=4$ super Yang-Mills \cite{Kim:2019feb,Kim:2019rwd}, and for mABJM theory \cite{Kim:2019ewv} which is obtained as a nontrivial fixed point of renormalization group after one of bi-fundamental chiral multiplet is given superpotential mass and integrated out. For the mass deformation of Brandhuber-Oz theory \cite{Jafferis:2012iv,Chang:2017mxc,Gutperle:2018axv,Brandhuber:1999np}, where numerical results did not lead to a definitive conjecture on analytic relations, we managed to obtain the holographic free energy in closed form, from the resummation of the perturbative results. This method is also successfully applied to supergravity Janus solutions \cite{Kim:2020unz} and a construction of de-Sitter solutions in massive IIA supergravity with O8-plane sources \cite{Kim:2020ysx}.

Going back to the main theme, in this paper we solve the HJ equation of the effective Lagrangian dual to ABJM model on $S^3$ with real mass terms. There are three pairs of real scalar fields, which were originally three complex scalars from three vector multiplets, interacting with a non-trivial potential. Our strategy is in the same spirit as the perturbative prescription mentioned in the last paragraph: Hamilton's characteristic function $W$ is expanded in scalar fields, and we show that the coefficients as a function of the warp factor satisfy an ordinary differential equation which can be integrated explicitly order by order. 

Then we illustrate how one can construct $W$ using the BPS equations. 
It turns out that, when the series-expansion ansatz of $W$ is substituted into the HJ-friendly version of BPS equations, $W$ can be determined {\it algebraically} from the series expansion form of the BPS equation. One does not have to solve differential equations any more. 

This paper is organized as follows. In Sec.\ref{Sec:2} we present the Einstein-scalar action, BPS equations, and the HJ equation. We explain why the Lagrangian is not reduced to a simple total-derivative for curved-slice, and how integrability is violated when the BPS equations are given in terms of $W$. In Sec.\ref{Sec:3} we construct $W$ in two different ways starting with the series-expansion form: first by solving HJ equation directly, and secondly using the BPS equations. We show how $W$ can be used to fix the coefficients of counterterms in holographic renormalization. Sec.\ref{Sec:4} is devoted to discussions.

\section{\label{Sec:2}Hamilton-Jacobi Approach to holography of ABJM with real mass}
\subsection{Setup and the Solution for AdS Vacuum}
We study the Einstein-scalar systems in four dimensions. This action is a truncation of STU supergravity \cite{Duff:1999gh}, which is in turn a consistent truncation of maximally supersymmetric $SO(8)$ gauged supergravity \cite{deWit:1982bul}. As we ignore the vector fields and axions, the action in Euclidean signature is written as follows \cite{Freedman:2013ryh}
\begin{align}
\label{action4}
S_{\rm sugra} &= \frac{1}{8\pi G_4} \int d^4 x \sqrt{g_4} \left[ -\frac{1}{2}R + \sum_{i=1}^3 \frac{\partial_\mu z_i \partial^\mu \tilde z_i}{(1-z_i \tilde z_i)^2} + \frac{1}{L^2} 
\left( 3 - \sum_{i=1}^3 \frac{2}{1-z_i \tilde z_i}\right)
\right]
.
\end{align}
There are originally four vector fields in STU supergravity, and they are dual to Cartan subalgebra of $SO(8)$ global symmetry in M2-brane theory. The three complex scalar fields (which become three pairs of real scalars in Euclidean signature) are dual to R-charge and real mass of matter fields in ABJM model.

For the spherically symmetric case, a metric choice which is most convenient for explicit integration of the equations is
\beq
ds^2 = e^{2A(r)}(dr^2/r^2 + ds^2_{S^3}) .
\eeq
Here $ds^2_{S^3}$ denotes metric of the round 3-sphere with unit radius.
For AdS (to be precise, the hyperbolic space as we are in Euclidean signature) vacuum, the scalars vanish and the warp factor is
\beq
e^{2A} = \frac{4r^2L^2}{(1-r^2)^2} .
\eeq

More generally the scalar fields are functions of $r$ and 
the associated BPS equations relevant to the above coordinate choice are given as follows.
\beq
\label{bps4}
r(1+  \tilde z_1 \tilde z_2 \tilde z_3 ) z_i^{\prime} &=
(\pm 1 -r A' )(1-z_i \tilde z_i )
\left( z_i + \frac{\tilde z_1 \tilde z_2 \tilde z_3}{\tilde z_i} \right) , 
\\
r(1+   z_1  z_2  z_3 ) \tilde z_i^{\prime} &=
(\mp 1 -r A' )(1-z_i \tilde z_i )
\left( \tilde z_i + \frac{z_1 z_2 z_3}{z_i} \right) ,
\\
-1 &= -r^2 (A^\prime)^2 +e^{2A}\frac{(1+z_1 z_2 z_3)(1+\tilde z_1 \tilde z_2 \tilde z_3)}{\prod_{i=1}^{3} 
(1-z_i \tilde z_i)} . 
\label{alsq}
\eeq
Here $(\bullet)':=\frac{d}{dr}(\bullet)$.
An exact solution, which is regular and with three integration constants dual to R-charge assignments of the matter fields in ABJM theory is presented in \cite{Freedman:2013ryh}
\beq
\label{solme}
e^{2A} &= \frac{4L^2r^2(1+c_1c_2c_3)(1+c_1c_2c_3 r^4)}{(1-r^2)^2(1+c_1c_2c_3r^2)^2} , 
\\
z_i(r) &= c_i f(r), \quad 
\tilde z_i (r) = -\frac{c_1c_2c_3}{c_i} f(r) ,
\\
\label{solsc}
f(r) &= \frac{1-r^2}{1+c_1c_2c_3 r^2} . 
\eeq
It will be refereed to as Freedman-Pufu (FP) solution.
Note that this solution can be also constructed by employing a perturbative method \cite{Kim:2019feb}, treating the integration constants $c_i$ as expansion parameters.

Let us now consider an effective Lagrangian obtained from the above gravity action by reducing on $S^3$. Using the following form of the metric
\beq
\label{mchoice}
ds^2 = d\tau^2 + \kappa^{-2} \alpha^2(\tau) ds^2_{S^3} , 
\eeq
the field equations are reduced to a coupled nonlinear differential equations. Here $\kappa$ is a parameter which will be set to $\kappa=g:=1/L$ later. For AdS vacuum, 
\beq
\label{vsol}
\alpha(\tau)= {\kappa}{g}^{-1} \sinh (g\tau) ,
\eeq
and the different choices of the ``radial'' coordinate are related through $r=\tanh (g\tau/2)$.

One can show that, if all scalar fields respect spherical symmetry, the field equations can be derived from the variation of the following mechanical Lagrangian with ``time'' $\tau$.
\begin{align}
\label{maction3}
    {\mathscr L} = 
    \alpha \dot\alpha^2 
    + {\kappa^2} \alpha 
     -\frac{\alpha^3}{3} \sum_{i=1}^3 \left[\frac{ { \dot z_{i}\dot{\tilde z}_{i}}}{(1-z_i \tilde z_i)^2} -{g^2}\frac{ 1+z_i\tilde z_i}{1- z_i \tilde z_i}\right]
     . 
\end{align}
One has to also augment it with the zero-energy condition, $\mathscr E=0$, which originates from the Hamiltonian constraint of Einstein's gravity.
This expression can be simply obtained by substituting the metric and the scalar field ansatz into the action \eqref{action4} with the Gibbons-Hawking term to remove the second derivative terms. Then we have identified $\int d\tau \mathscr L = - 8\pi G_4 S_{\rm sugra}$. 
Also note that although setting $\kappa=0$ in the metric ansatz \eqref{mchoice} and the AdS vacuum solution \eqref{vsol} look singular, it is just the limit where the domain wall has a flat slice.

A comment is in order now on how the Lagrangian simplifies on-shell. It is a well-known procedure in soliton physics that by completing the square of the Lagrangian one can sometimes derive BPS equations and the Lagrangian is reduced to a total derivative, which should be equal to Hamilton's principal function. But it is not the case in general, especially for the class of mechanical systems we are currently interested in, which is derived from general relativity action with a curved slice. 

In fact, when we substitute the equation of motion for $\alpha$ into the action \eqref{maction3}, we find 
\beq
\mathscr L_{\rm on-shell} = \frac{2}{3} \frac{d}{d\tau} \left( \alpha^2 \dot \alpha \right) + \frac{2}{3}\kappa^2\alpha .
\eeq
We thus see that $\mathscr L$ would be total derivative, if we considered a flat slice ansatz {\it i.e.} $\kappa\ra0$. It does not improve even if we make use of BPS conditions, since $\alpha=\kappa e^A$ is {\it not} a total derivative, as one can infer from \eqref{alsq}. 

We note that this happens when the curved slice is more than two-dimensional. If we considered an analogous problem in three-dimensional gravity with $S^2$ or $AdS_2$ slices, the last term would be absent and the action would be a total-derivative. It also follows that, when one considers non-rotating, spherically symmetric black hole solutions in four-dimensional gravity the Lagrangian is total-derivative on-shell. For recent works which exploit this property, readers are referred to \cite{Cabo-Bizet:2017xdr,Bobev:2020pjk}.

\subsection{Single-scalar Reduction}
In order to convey the essence of our procedure, let us now restrict ourselves to a simple subclass of solutions where we keep only one pair scalars. Solutions to the original model will be presented in Sec.\ref{SSec:3.3}. 

We set $z:=z_1=z_2=z_3$ and $\tilde z:=\tilde z_1=\tilde z_2=\tilde z_3$ and we call this simplified model a single-scalar case because it is originally a reduction to a single complex-scalar model. On the dual field theory side, this means that we give the same R-charge (real mass) to three chiral multiplets out of four. Now we have a single-scalar Lagrangian,
\begin{align}
    \mathscr L =  \alpha \dot\alpha^2 - \frac{\alpha^3 { \dot z\dot{\tilde z}}}{(1-z \tilde z)^2}
    + \kappa^2 \alpha + g^2\alpha^3 \frac{1+z\tilde z}{1- z \tilde z} . 
\end{align}

Through Legendre transformation and considering a canonical transformation to a trivial Hamiltonian, one obtains the Hamilton-Jacobi equation for Hamilton's principal function $S(\alpha,z,\tilde z;\tau)$.
\beq
-\frac{\partial S}{\partial \tau}
= \frac{1}{4\alpha}\left(\frac{\partial S}{\partial \alpha}\right)^2 
- \frac{(1- z \tilde z)^2}{\alpha^3} \frac{\partial S}{\partial z} \frac{\partial S}{\partial \tilde z} 
 - \kappa^2 \alpha-g^2 \alpha^3\frac{1+z \tilde z}{1- z \tilde z} . 
\eeq
We note that usually in the HJ formalism of holographic renormalization one utilizes the ADM formalism and retain the general covariance on boundary \cite{Papadimitriou:2004ap}, but for simplicity here we consider the dimensionally reduced Lagrangian given above.

Since the Hamiltonian has no explicit ``time'' $(\tau)$ dependence, we can employ the usual additive separation of variables and write
\beq
S (\alpha,z,\tilde z;\tau)= W(\alpha,z,\tilde z) - \mathscr E \tau ,
\eeq
where $W$ is Hamilton's characteristic function. It satisfies
\beq
\frac{1}{4\alpha}\left(\frac{\partial W}{\partial \alpha}\right)^2
-\frac{(1- z \tilde z)^2}{\alpha^3} \frac{\partial W}{\partial z} \frac{\partial W}{\partial \tilde z} 
  - \kappa^2 \alpha-g^2 \alpha^3\frac{1+z \tilde z}{1- z \tilde z} = \mathscr E . 
\label{hj1}
\eeq

Recall that physical solutions should satisfy $\mathscr E=0$, but in order to obtain $\tau$ dependence of the degrees of freedom, we need to temporarily consider $\mathscr E\neq 0$. Since $\mathscr E$ is one of new canonical momenta and its conjugate is also a constant of motion, 
\beq
\frac{\partial W}{\partial \mathscr E} = \tau .
\eeq
We can easily check this leads to the AdS vacuum solution when the scalars are set to zero. Let us call the characteristic function for the vacuum $W_0$. We find that the Hamilton-Jacobi equation \eqref{hj1} becomes
\beq
\frac{\partial W_0}{\partial \alpha} = \pm \sqrt{ 4\alpha (\kappa^2\alpha+g^2\alpha^3+\mathscr E)} . 
\eeq
The sign ambiguity reflects invariance under $\tau\ra -\tau$. With a positive sign, we first expand the integrand up to linear order in $\mathscr E$ and perform integration
\beq
\label{w0g}
W_0 = \frac{2}{3g^2} (\kappa^2+g^2\alpha^2)^{3/2} + \frac{\mathscr E}{g} \sinh^{-1} \left(
\frac{g\alpha}{\kappa} \right) + {\cal O}(\mathscr E^2).
\eeq

For the solution satisfying $\mathscr E=0$, we have the following results which agree with $\alpha={\kappa}{g}^{-1} \sinh (g\tau)$.
\beq
\tau&=\left.\frac{\partial W_0}{\partial \mathscr E} \right|_{\mathscr E=0} = \frac{1}{g} \sinh^{-1}  \left(
\frac{g\alpha}{\kappa} \right) ,
\\
\frac{\partial \mathscr  L}{\partial \dot \alpha} =2\alpha \dot \alpha 
&= \left.\frac{\partial W_0}{\partial \alpha}\right|_{\mathscr E=0} = 2\alpha \sqrt{\kappa^2+g^2\alpha^2} .
\eeq

Because our aim is to obtain the characteristic function $W$ as the evaluated action for FP solution, we will restrict ourselves to $\mathscr E=0$ from now on for simplicity. It means that we have only the second equation in the above, which can be integrated to give the first equation. When we turn on $z,\tz$ we have coupled differential equations which do not readily give $\tau$ dependence. Although $\tau$ dependence cannot be obtained right away, One can still check whether a given $W$ describes the FP solution or not, through the implicit relations between $\alpha,z,\tilde z$ and their conjugate momenta from $W$.

\subsection{Non-integrability of BPS equations}
Before we derive the solutions for $W$, let us discuss why it is a non-trivial problem even for supersymmetric solutions, when we consider curved slices ($\kappa\neq 0$). The BPS equation for the single-scalar model in the metric choice \eqref{mchoice} becomes
\begin{align}
\label{bps1}
    \dot \alpha^2 &= \kappa^2  + g^2\alpha^2\frac{(1+z^3)(1+\tilde z^3)}{(1-z \tilde z)^3} , 
    \\
   \alpha \dot z&= (\pm \kappa  - \dot \alpha )\frac{(1-z \tilde z)\left( z + \tilde z^2 \right)}{1+\tilde z^3} ,
    \\
\label{bps3}
\alpha\dot   { \tilde z} &= (\mp \kappa - \dot \alpha)\frac{(1-z \tilde z)\left( \tilde z + z^2 \right)}{1+z^3} .
\end{align}

These first-order relations are enough to guarantee that the equations of motion are all satisfied. One can also easily check that substituting these equations into the Hamiltonian, it leads to $\mathscr H=0$. In order to see the relation between the explicit solutions presented above in \eqref{solme}-\eqref{solsc} and the results we will obtain in our gauge choice with $\tau$ here, we choose the upper choice of signs in \eqref{bps1}-\eqref{bps3}, set $c:=c_1=c_2=c_3$, and recall that there is a relation between two different parametrizations,
\begin{align}
    \frac{d\tau}{dr} = \frac{2\sqrt{1+c^3}\sqrt{1+c^3r^4}}{g(1-r^2)(1+c^3r^2)} . 
\end{align}

The BPS equations \eqref{bps4}-\eqref{alsq} are derived from the supersymmetry condition, {\it i.e.} by setting the variation of fermion fields to zero and looking for non-trivial Killing spinors and the associated projection rules. The supersymmetry transformation rules are given in terms of the K\"ahler potential and the prepotential of $\cN=2$, $D=4$ supergravity. It follows that the right-hand-side expressions in \eqref{bps1}-\eqref{bps3}, for the mechanical model dimensionally reduced from $D=4$ gravity, are also summarized in terms a superpotential which is closely related to the aforementioned data of the supergravity. 

We find that the most convenient choice is 
\begin{align}
\label{sp}
    {\cal W}_0 =  \frac{2g\alpha^{3}}{3} \sqrt{\frac{(1+z^3)(1+\tilde z^3)}{(1-z \tilde z)^3}} . 
\end{align}
Then it is easy to see that it is also a solution of the characteristic function $W$, when $\kappa=0$. Namely, one can check $\left.{\partial \mathscr L}/{\partial \dot\alpha}\right|_{\kappa=0} = {\partial \cW_0}/{\partial \alpha}$ and also $\left.{\partial \mathscr L}/{\partial \dot z}\right|_{\kappa=0} = {\partial \cW_0}/{\partial  z}$, $\left.{\partial \mathscr L}/{\partial  \dot\tz}\right|_{\kappa=0} = {\partial \cW_0}/{\partial \tz}$. We note that this expression is proportional to $\sqrt{h}e^{{\cal K}/2}|W_{\rm sugra}|$, where $h$ is induced metric on the boundary $S^3$, ${\cal K}$ is the K\"ahler potential, $W_{\rm sugra}$ is the superpotential of $\cN=2$ supergravity. And this is also the supersymmetric counterterm introduced in \cite{Freedman:2013ryh}, specialized to the $SU(3)$ symmetric case we consider here. 

On the other hand, when $\kappa\neq 0$ we do not enjoy such a nice property any more. The BPS equations \eqref{bps1}-\eqref{bps3} can be re-written in a HJ-friendly form,
\beq
\label{bpsal}
\frac{\partial W}{\partial\alpha} &=2\alpha
\sqrt{\kappa^2  + g^2\alpha^2\frac{(1+z^3)(1+\tilde z^3)}{(1-z \tilde z)^3}} =
 \sqrt{4\kappa^2 \alpha^2 + \left(\frac{\partial\cW_0}{\partial\alpha}\right)^2},
\\
\frac{\partial W}{\partial \tz}&=
\left( \sqrt{\kappa^2+ g^2\alpha^2\frac{(1+z^3)(1+\tilde z^3)}{(1-z\tilde z)^3} } \mp \kappa
\right)
\frac{\alpha^2\left( z + \tilde z^2 \right)}{(1-z \tilde z)(1+\tilde z^3)}
\label{bpstz}
\nn\\
&= \left( \frac{\partial W}{\partial \alpha} \mp {2\kappa\alpha}\right) \frac{\alpha}{3\cW_0}\frac{\partial \cW_0}{\partial \tz},
\\
\frac{\partial W}{\partial z}&=
\left( \sqrt{\kappa^2+ g^2\alpha^2\frac{(1+z^3)(1+\tilde z^3)}{(1-z\tilde z)^3} } \pm \kappa
\right)
\frac{\alpha^2\left( \tilde z + z^2 \right)}{(1-z \tilde z)(1+ z^3)}
\nn\\
&= \left( \frac{\partial W}{\partial \alpha} \pm {2\kappa\alpha}\right) \frac{\alpha}{3\cW_0}\frac{\partial \cW_0}{\partial z} .
\eeq
We see that $W=\cW_0$ is obviously a solution when $\kappa=0$, if we recall $\alpha\frac{\partial \cW_0}{\partial \alpha} = 3\cW_0$. When $\kappa\neq 0$, apparently such an easy integration is not available, even though these equations certainly hold when we substitute the solutions $\alpha(\tau),z(\tau),\tz(\tau)$. In fact, there is an obstruction: A necessary condition for integrability of $W(\alpha,z,\tz)$ is $\frac{\partial}{\partial z}\frac{\partial W}{ \partial \tz} = \frac{\partial}{\partial \tz}\frac{\partial W}{ \partial z} $ and so on. Acting on the right-hand-side expressions, they are not identical as functions of $\alpha,z,\tz$. Let us re-cap what we have discovered. In the HJ approach, the BPS equations are not readily integrable, and they hold only {\it on-shell}, {\it i.e.} after we substitute the solutions as functions of $\tau$.

One might wonder why turning on $\kappa$ makes such a big difference. In fact, already with the AdS vacuum, having $S^3$ instead of $\mathbb{R}^3$ incurs a big difference with bulk (IR) behavior. It is seen from the form of AdS solution $\alpha = \kappa/g \sinh(g\tau)$, and also from the potential part of the Lagrangian $\mathscr L$ when $\alpha$ is small. It is also understandable from the AdS/CFT point of view, since by putting the theory on Euclidean sphere we are introducing an IR cutoff, set by the radius of $S^3$.

\section{\label{Sec:3}Solutions to Hamilton's Characteristic Function}
\subsection{Solutions of Single-Scalar Model}
Having stressed the non-triviality of the problem, we provide a recursive method of solving the HJ equation. From now on we set $\kappa=g$ to simplify the formulae.
Let us consider a re-parametrization
\begin{align}
    x = z \tilde z, \quad y = z / \tilde z ,
\end{align}
which implies
\begin{align}
\dot z \dot \tz &= \frac{y^2{\dot x^2} - x^2 {\dot y^2}}{4xy^2}  , 
\\
\frac{\partial W}{\partial z} \frac{\partial W}{\partial \tilde z}
&= x \left( \frac{\partial W}{\partial x} \right)^2 - \frac{y^2}{x} \left( \frac{\partial W}{\partial y} \right)^2 . 
\end{align}
Then the HJ equation is written as 
\begin{align}
    \frac{1}{4\alpha}\left(\frac{\partial W}{\partial \alpha}\right)^2 -\frac{(1- x)^2}{\alpha^3x} \left[ x^2 \left( \frac{\partial W}{\partial x} \right)^2 - {y^2} \left( \frac{\partial W}{\partial y} \right)^2 \right]    - g^2 \alpha-g^2 \alpha^3\frac{1+x}{1- x}
=0 .
\end{align}

Unfortunately, because of the last term we cannot solve this equation generally using separation of variables. In order to make a progress, we restrict ourselves to the case where $W$ is independent of $y$. In fact this is consistent with the special property of the FP solution, where $y=z/{\tilde z}$ is constant, implying $\partial W/\partial y = 0$.  Our strategy is to express $W$ as a series expansion of $x$, where the coefficients are functions of $\alpha$.
\begin{align}
\label{wexpansion}
    W (\alpha, x)= -\frac{2g}{3} + W_0(\alpha) \left( 1 + \sum_{n=1}^\infty (1+\alpha^2)^{-n} w_n (\alpha) x^n \right) .
\end{align}
We have added a constant $-2g/3$, which of course makes no difference to the HJ equation, in order to impose $W|_{\alpha=0}=0$. It is because we intend to compare $W$ with the supergravity action evaluated using the BPS solutions.
Putting a factor of $(1+\alpha^2)^{-n}$ is not essential but we find empirically that it simplifies the differential equation for $w_n$. Then each of $w_n$, except for $w_1$, can be determined by solving a linear ordinary differential equation, whose coefficients are given by lower-index $w_n$'s. For $w_1$, as we will see, we obtain a non-linear differential equation which can be fortunately solved exactly.

We already know from \eqref{w0g}
\begin{align}
    W_0 = \frac{2g}{3} (1+\alpha^2)^{3/2} . 
\end{align}
From the terms linear in $x$, we find that $w_1$ should satisfy
\begin{align}
\label{w1eq}
     3\alpha ^3\left(\alpha
   ^2+1\right)  w_1' -2 (1+\alpha^2) w_1^2 + 3 \alpha ^4
   w_1 = 9 \alpha^6 . 
\end{align}
This can be integrated in general,
\begin{align}
\label{w1res}
    w_1 &= \frac{3\alpha^2\left( 2+\alpha^2+2c_1 \sqrt{1+\alpha^2}\right)}{2(1+\alpha^2 + c_1 \sqrt{1+\alpha^2})} .
\end{align}

The correct value of the integration constant $c_1$ is fixed by demanding regularity at IR, {\it i.e.} $\alpha=0$. The above expression is $\cO(\alpha^2)$ for generic values of $c_1$, then the rate at which $S^3$ shrinks gets changed and a conical singularity is developed. Only if $c_1=-1$, $w_1\sim \cO(\alpha^4)$ and the solution remains regular at IR. The answer is then
\begin{align}
    w_1 = \frac{3\alpha^2}{2} \left( 1 - \frac{1}{\sqrt{1+\alpha^2}} \right) .
    \label{w1}
\end{align}

Let us now check if this result is consistent with the FP solution. There are various things one can compare, but let us see if the on-shell value of the action agrees. The approximation to $\cO(x)$ amounts to keeping up to $\cO(c^3)$ terms in the FP solution, which in the single-scalar model becomes
\begin{align}
    \alpha & = \frac{2r\sqrt{1+c^3}\sqrt{1+c^3r^4}}{(1-r^2)(1+c^3r^2)} , 
    \\
    x & = - \frac{c^3(1-r^2)^2}{(1+c^3r^2)^2} . 
\end{align}

When we recall $\mathscr H = 0$, the on-shell value of the Lagrangian becomes minus two times the potential part, so the action is calculated as follows
\beq
S &=
\int_0^r 2g^2 \alpha \left( 1 + \alpha^2 \frac{1+x}{1-x} \right)  \frac{d\tau}{dr} \, dr
    \nn\\
   &= \frac{4gr^2(3+r^4)}{3(1-r^2)^3}  +4gc^3r^2 +\cO(c^6) .
   \label{action1}
\eeq
To obtain the second line above we substitute the solutions $\alpha(r),x(r)$ into the integrand, perform the integration, and keep terms only up to $\cO(c^3)$. On the other hand, one can evaluate $W$ in \eqref{wexpansion} up to linear order in $x$ by substituting the FP solution. The result matches exactly with \eqref{action1}.

At order of $x^2$, we obtain a first-order {\it linear} differential equation for $w_1$ which can be explicitly integrated. Again, by demanding it should be $\cO(\alpha^4)$ for small $\alpha$, we can fix the integration constant. The result is 
\begin{align}
    w_2=\frac{3}{8}\alpha^2 \left(\sqrt{\alpha^2+1}-1\right)\left(4 \sqrt{\alpha^2+1}-\alpha^2\right) . 
\end{align}
After that one repeats the same procedure. We only provide the results for $w_3,w_4$ below.
\begin{align}
    w_3&=\frac{1}{16} \alpha^2\left(\sqrt{\alpha^2+1}-1\right)\left[(\alpha^4+24\alpha^2+24)\sqrt{\alpha^2+1}+\alpha^6-10 \alpha^4 -12\alpha^2\right],
\\
     w_4&=-\frac{3}{128}  \alpha^2 \left(\sqrt{\alpha^2+1}-1\right)\left[ \left(\alpha^8-6 \alpha^6-72 \alpha^4-128 \alpha^2-64\right)\sqrt{\alpha^2+1}\right.
     \nn\\
     &+\alpha^{10}-5 \alpha^8+27 \alpha^6+80 \alpha^4+48 \alpha^2\Big].
\end{align}
They include irrational functions in $\alpha$, but if we choose to rewrite in terms of {\it e.g.} $\gamma^2={\sqrt{1+\alpha^2}-1}$, they can be written as a polynomial.

Although it looks unlikely that the series \eqref{wexpansion} in $x$ with $w_n$ as the coefficients can be re-summed, one could have obtained the same result, from the on-shell action \eqref{action1}. It can be calculated, using the FP solution, exactly as a function of $r$ and $c$. At the same time, $\alpha$ and $x=z\tz$ are also functions of $r$ and $c$. If one inverts these relations and substitute into \eqref{action1}, we would obtain $W(\alpha,x)$. Let us also comment that such a brute-force derivation may work here, because there is an integration constant $c_i$ for each variable $x_i:=z_i\tz_i$.  

Let us now study the divergence of $W$ as $\alpha\ra\infty$, which corresponds to UV limit in the context of holography. This will tell us what kind of counterterms on the boundary should be added for holographic renormalization. By analyzing the equations of $\dot\alpha$ and $\dot x$ from derivatives of $W$, one can argue that when $\alpha\ra\infty$, $x$ vanishes just as in the FP solution, keeping $\alpha^2 x$  constant. 

Then, from the UV behavior of $w_n$, we see that $W_0$ part shows cubic divergence, the part linear in $x$ shows linear divergence, while higher-order terms give finite contributions in UV. More concretely, in the UV
\beq
W = \frac{2g}{3} \left( \alpha^3 + \frac{3}{2}\alpha + \frac{3}{2} \alpha^3 x
 \right) + \mbox{finite terms}.
\eeq
The first and the third terms can be simultaneously cancelled by a {\it supersymmetric} counterterm $-\cW_0$, because 
\beq
S_{susy} := -\cW_0 = -\frac{2g\alpha^{3}}{3} \sqrt{\frac{(1+z^3)(1+\tilde z^3)}{(1-z \tilde z)^3}} = -\frac{2g}{3} \alpha^3 \left( 1+ \frac{3}{2} x \right) + \mbox{finite terms} . 
\eeq
And the second term with linear divergence can be removed by adding the {\it boundary curvature} term, $\sqrt{h} {\cal R}$ with an appropriate choice of the coefficient. Here $h$ is the induced metric, and ${\cal R}$ is the scalar curvature on the boundary $S^3$.
Then one can easily verify that the remaining finite pieces give exactly the same result as Eq.(6.19) in \cite{Freedman:2013ryh}.
\subsection{Characteristic function from BPS equations}
In the last subsection we have obtained a solution to HJ equation and argued that it corresponds to the FP solution. It is rather surprising that we managed to obtain $W$ for supersymmetric solutions, when we did not explicitly make use of the BPS equations. Our success implies that the assumption of ${\partial W}/{\partial y}=0$, which means $z/\tz$ is constant {\it i.e.} $z$ and $\tz$ have the same profile, is strong enough to guarantee supersymmetry, when combined with the equations of motion. We note that at every order in $x$ one is required to solve a first-order ordinary differential equation.

We have already pointed out that, unlike the flat slice case, it is not feasible to integrate the BPS equations immediately  to obtain the characteristic function. However, having obtained the solution iteratively as a series expansion in $x=z\tz$, in this subsection we explain how BPS equations can be indeed used to find $W$, again order by order in $x$. A notable and advantageous feature of this procedure is that we do not need to solve differential equations, and the computation needed is just series-expansion. 

Let us write 
\beq
W(\alpha,x) = \sum_{n=0}^\infty W_n(\alpha)x^n .
\eeq
We know $W_0$ can be easily integrated. 
We then consider the BPS equations for $\dot z, \dot \tz$ and expand it to the leading non-trivial order. Although we know $z\sim \cO(c^3)$ and $\tz\sim \cO(c^6)$ from explicit solutions, let us pretend we do not know this fact yet. Assuming (incorrectly) $z,\tz$ are of the same order, from the BPS equations we have the following conditions
\beq
\frac{\partial W}{\partial \tz} & = W_1' z + \cdots = g\alpha^2 (\sqrt{1+\alpha^2}-1) z +\cdots,
\label{z1}
\\
\frac{\partial W}{\partial z} & = W_1' \tz + \cdots= g\alpha^2 (\sqrt{1+\alpha^2}+1) \tz + \cdots.
\label{tz1}
\eeq
We clearly see a problem, since these two equations are in contradiction with each other. At least one of them must be wrong, and we would like to devise a general procedure, by which one can fix the error and construct $W$.

We recall that, when we solve the BPS equations as a differential equation, and also when solving the HJ equation in the last subsection, IR regularity is an important guideline. As we consider small $\alpha$ limit, $z$ and $\tz$ are non-zero but $\dot z$ and $\dot \tz$ should vanish. Then, obviously \eqref{tz1} is the wrong one, and let us keep \eqref{z1} as the correct one. 
\beq
W_1 = g\alpha^2(\sqrt{1+\alpha^2}-1) . 
\eeq
This of course agrees exactly with the first-order result we obtained in \eqref{w1}.

Before we try to fix the problem in \eqref{tz1}, 
let us consider the BPS equation for $\dot\alpha$, \eqref{bpsal}. Since we know $z$ and $\tz$ should not be treated as of the same order, we assume (correctly) $\tz$ and $z^2$ are of the same order. Then we obtain
\beq
W_1' z \tz = g\alpha^3 \frac{z^3+3z\tz}{\sqrt{1+\alpha^2}} .
\eeq
Because we have already obtained $W_1$, this condition gives us an {\it on-shell} relation between $z$ and $\tz$.
\beq
\label{ac1}
\alpha^2z^2 = 2({1-\sqrt{1+\alpha^2}}) \tz.
\eeq
One can easily check it is consistent with FP solution, when it is truncated at $\cO(c^6)$.

We can now fix the trouble with \eqref{tz1}. We include terms of order $z^2$, and make use of the on-shell relation above. 
\beq
\frac{\partial W}{\partial z} & = g\alpha^2 (\sqrt{1+\alpha^2}+1) (\tilde z + z^2) + \cdots
\\
 &= g\alpha^2 (\sqrt{1+\alpha^2}+1)
\left( 1 + \frac{2({1-\sqrt{1+\alpha^2}})}{\alpha^2}\right) \tilde z +\cdots
\nn\\
& = g\alpha^2 (\sqrt{1+\alpha^2}-1) \tilde z . 
\eeq
We thus see it also leads to $W_1 = g\alpha^2(\sqrt{1+\alpha^2}-1)$, and everything fits together.

In general, we can proceed repeating the expansion, assigning a weight 1 and 2 on $z$ and $\tz$ respectively. We turn to ${\partial W}/{\partial \tz}$ again, now keeping up to terms of weight 4.
\beq
\frac{\partial W}{\partial \tilde z} 
&=
g\alpha^2 \left\{ \frac{\alpha^2(z^3+3z\tilde z)}{2\sqrt{1+\alpha^2}} +
(\sqrt{1+\alpha^2}-1)\tilde z \right\} z^2+g\alpha^2 (\sqrt{1+\alpha^2}-1)\tilde z^2
+\cdots. 
\eeq
If one then makes use of the on-shell relation \eqref{ac1}, the expression can be rewritten as
\beq
\frac{\partial W}{\partial \tilde z} = W_1(\alpha) z + 2 W_2(\alpha) z^2 \tz + \cdots,
\eeq
where $W_2$ agrees with the result obtained by integrating the HJ equation. We have verified this procedure can be applied repeatedly and obtained the same results as in the last sub-section.
\subsection{\label{SSec:3.3}Multi-Scalar solutions}
So far we have restricted ourselves to the single-scalar model, just for simplicity. It should be obvious now that our method can be applied to the original three-scalar model as well. In this sub-section we present the essential steps and the result.

The Hamilton-Jacobi equation for characteristic function is now
\beq
\frac{1}{4\alpha}\left(\frac{\partial W}{\partial \alpha}\right)^2 - g^2 \alpha
-\frac{1}{3}\sum_{i=1}^3 \left[\frac{(1- z_i \tilde z_i)^2}{\alpha^3} \frac{\partial W}{\partial z_i} \frac{\partial W}{\partial \tilde z_i} 
  +g^2 \alpha^3\frac{1+z_i \tilde z_i}{1- z_i \tilde z_i}
  \right] = 0 . 
\label{hj3}
\eeq

Due to symmetry, the solution can be expanded as follows in terms of $x_i:=z_i\tz_i$, and a symmetry argument restricts $W$ as follows.
\begin{align}
    W(\alpha,x_i) &= \sW_0 (\alpha) + \sW_1(\alpha) (\tfrac{1}{3}\sum x_i) 
    \nn\\
    & + \sW_2^{(1)}(\alpha) (\tfrac{1}{3}\sum x_i^2)
    + \sW_2^{(2)}(\alpha) (\tfrac{1}{3}\sum x_i)^2
     \nn\\
    & + \sW_3^{(1)}(\alpha) (\tfrac{1}{3}\sum x_i^3)
    + \sW_3^{(2)}(\alpha) (\tfrac{1}{3}\sum x_i^2)(\tfrac{1}{3}\sum x_i)
    + \sW_3^{(3)}(\alpha) (\tfrac{1}{3}\sum x_i)^3
    \nn\\
    &+\sW_4^{(1)} (\alpha)(\tfrac{1}{3}\sum x_i^4)+\sW_4^{(2)}(\alpha) (\tfrac{1}{3}\sum x_i^3) (\tfrac{1}{3}\sum x_i)
    \nn\\
    &+\sW_4^{(3)}(\alpha) (\tfrac{1}{3}\sum x_i^2)^2 +\sW_4^{(4)}(\alpha) (\tfrac{1}{3}\sum x_i^2)(\tfrac{1}{3}\sum x_i)^2
    +\cdots . 
\end{align}

We give the results below, expressed in terms of $\gamma$ defined through $\alpha^2=2\gamma^2+\gamma^4$ or equivalently $\gamma^2=\sqrt{1+\alpha^2}-1$.
\begin{align}
    \sW_0 & =W_0 
    = \frac{2}{3}g (1+\gamma^2)^{3} . 
\end{align}
\begin{align}
    \sW_1
    &=g \gamma^4 (2+\gamma^2) .
\end{align}
\begin{align}
    \sW_2^{(1)}
    &=\frac{g\gamma^4(2+\gamma^2)}{6(1+\gamma^2)}(6+ \gamma^2) . 
\end{align}
\begin{align}
    \sW_2^{(2)} 
    &=-\frac{g}{12(1+\gamma^2)}\gamma^6(2+\gamma^2)\left(8+5\gamma^2\right).
\end{align}
\begin{align}
    \sW_3^{(1)}
    &=\frac{g\gamma^4(2+\gamma^2)}{360}(360+120\gamma^2+30 \gamma^4 +7 \gamma^6) . 
\end{align}
\begin{align}
    \sW_3^{(2)}
    &=-\frac{g\gamma^6(2+\gamma^2)}{72(1+\gamma^2)}(96+102\gamma^2+37\gamma^4+7\gamma^6) . 
\end{align}
\begin{align}
    \sW_3^{(3)}
    &=\frac{g\gamma^8(2+ \gamma^2)}{360(1+\gamma^2)^3}(360+868\gamma^2+789\gamma^4+309\gamma^6+43\gamma^8) . 
\end{align}
\begin{align}
    \sW_4^{(1)}
    &=-\frac{g\gamma^4(2+\gamma^2)}{38880(1+\gamma^2)^5}(-38880-213840\gamma^2-495720\gamma^4-619164\gamma^6-435974\gamma^8
    \nn\\
    &-145109\gamma^{10}+21734\gamma^{12}+47796\gamma^{14}+23200\gamma^{16}+5282\gamma^{18}+470\gamma^{20}) . 
\end{align}
\begin{align}
    \sW_4^{(2)}
    &=\frac{g\gamma^6(2+\gamma^2)}{136080(1+\gamma^2)^5}(-181440-997920\gamma^2-2137212\gamma^4-2225482\gamma^6
    \nn\\
    &-945163\gamma^8+280039\gamma^{10}+528186\gamma^{12}+260072\gamma^{14}+59539\gamma^{16}+5311\gamma^{18}) . 
\end{align}
\begin{align}
    \sW_{4}^{(3)}
    &=\frac{g\gamma^6(2+\gamma^2)}{181440(1+\gamma^2)^5}(-120960-665280\gamma^2-1449504\gamma^4-1575224\gamma^6
    \nn\\
    &-778436\gamma^8+51683\gamma^{10}+277302\gamma^{12}+147844\gamma^{14}+34673\gamma^{16}+3107\gamma^{18}) . 
\end{align}
\begin{align}
    \sW_4^{(4)}    &=-\frac{g\gamma^8(2+\gamma^2)}{30240(1+\gamma^2)^5}(-90720-337008\gamma^2-484148\gamma^4-273218\gamma^6
    \nn\\
    &+54413\gamma^8+157602\gamma^{10}+85180\gamma^{12}+20099\gamma^{14}+1805\gamma^{16}) . 
\end{align}
\section{\label{Sec:4}Discussion}
We have studied the Hamilton-Jacobi (HJ) approach to the BPS equations associated with the holography of ABJM model with real mass terms/R-charge assignments. The field theory side computation can be done using the supersymmetric localization technique, and when the theory is put on $S^3$ taking the large-$N$ limit and extracting the leading-order $N^{3/2}$ behavior from the matrix model with the correct dependence on R-charges is straightforward. 

On the holography side the explicit solutions found by Freedman and Pufu makes the analysis rather easy, but how to solve the HJ equation to obtain the on-shell action directly remained unaddressed. We find it appropriate to start the HJ approach with FP's BPS equations to establish the prescription, and then try to apply more challenging problems like $\cN=1^*$ or mass-deformed Brandhuber-Oz theory in the future.

We have constructed the solutions to HJ equation in two ways, first without relying on BPS equation, and then again using the BPS conditions. In particular, we find it satisfactory that the BPS equations allow one to find the characteristic function $W$ through only algebraic manipulations, without having to solve a differential equation. 

Readers might be puzzled by the fact that our solutions do not include any integration constant. It is just a function of $\alpha$ and $x$, and in particular does not include the parameter $c$ which controls the IR value of the scalar fields. It is because $c$ gives the value of $y=z/\tz=-c^{-1}$ which is constant. It is akin to the fact that for a free particle at rest the principal function is simply $S=0$ and the position of the particle  does not appear. What we can use to express the renormalized supergravity action is $\lim_{\alpha\ra\infty}(\alpha^2 z\tz)$ instead.

Finally, it is natural to ask whether HJ approach will prove powerful with other examples where explicit solutions are not available and the evaluation of on-shell action remains a conjecture. The easiest next problem to tackle is probably mass-deformed Brandhuber-Oz theory  \cite{Gutperle:2018axv} where the holographic free energy is known \cite{Kim:2019feb}, although explicit solutions to BPS equation is not available. The most intriguing problem is of course mass deformation of $\cN=4$ super Yang-Mills. For $\cN=2^*$ we are given a conjecture \cite{Bobev:2013cja}, which is yet to be proved rigorously, on the holographic free energy formula which is consistent with localization results. For $\cN=1^*$ \cite{Bobev:2016nua}, localization is not applicable and holographic computation has been done only up to fifth order using the perturbation method \cite{Kim:2019rwd}. It will be interesting to see if HJ approach reveals new insights into these problems.  
\section*{Acknowledgements}
We thank Kimyeong Lee for drawing our attention to Ref.\cite{Kim:2007aa}, and I. Papadimitriou for valuable comments and a summary of his works on HJ approach to AdS/CFT.
This work was done partly during the 24th APCTP Winter School on Fundamental Physics, Jan. 30 -- Feb. 5 2020, and we appreciate the hospitality. This research was supported by the National Research Foundation (NRF) grant 2019R1A2C2004880. 
\bibliographystyle{JHEP}
\bibliography{ref}
\end{document}